# Mapping ChatGPT in Mainstream Media to Unravel Jobs and Diversity Challenges:

Early Quantitative Insights through Sentiment Analysis and Word Frequency Analysis

Maya Karanouh
mkaranouh@brocku.ca
PhD Program of Interdisciplinary Humanities, Brock University, Canada

*Abstract -* The exponential growth in user acquisition and popularity of OpenAI's ChatGPT, an artificial intelligence(AI) powered chatbot, was accompanied by widespread mainstream media coverage. This article presents a quantitative data analysis of the early trends and sentiments revealed by conducting text mining and NLP methods onto a corpus of 10,902 mainstream news headlines related to the subject of ChatGPT and artificial intelligence, from the launch of ChatGPT in November 2022 to March 2023. The findings revealed in sentiment analysis, ChatGPT and artificial intelligence, were perceived more positively than negatively in the mainstream media. In regards to word frequency results, over sixty-five percent of the top frequency words were focused on Big Tech issues and actors while topics such as jobs, diversity, ethics, copyright, gender and women were poorly represented or completely absent and only accounted for six percent of the total corpus. This article is a critical analysis into the power structures and collusions between Big Tech and Big Media in their hegemonic exclusion of the "Other" from mainstream media.

I. INTRODUCTION

In November 2022, OpenAI launched ChatGPT 3, their artificial intelligence(AI) powered chatbot to the public. ChatGPT 3, a language processing AI model, was trained on 175 billion parameters with Big Data sets of over 300 billion words obtained from the internet, Wikipedia and other text data found on the internet (Hughes 2023). In the first week of ChatGPT's release over one million users logged in to interact with the newly released chatbot and it currently has one of the fastest-growing userbases, reaching 100,000 million users in two months (Sier 2022; Hu 2023). To frame this in terms of scale of other popular social media platform's user acquisition time frames, Instagram took two and half years and TikTok took nine months to reach the same goal of 100,000 million users (Hu 2023). This exponential growth in user acquisition and the rise of public interest in ChatGPT and artificial intelligence was reflected in the direct increase in the publication of news articles in mainstream media about ChatGPT and artificial intelligence.

In their book Media/Society: Industries, Images, and Audiences, David Croteau and William Hoynes argue that in our contemporary world, mainstream media has become the dominant influencer of society, surpassing older influential systems such as educational and religious institutions (2013). Mass media has the ability to focus attention to certain topics, and influence the public perception in what is known as agenda-setting theory (McCombs & Reynolds 2002). According to agenda-setting theory, the media is effective in "shaping and manipulating" the public's preferences on political, economic, and social issues. It also has the power to draw the public's attention to particular "events, issues and persons and in determining the importance people attach" to them (McCombs & Reynolds 2002). In mainstream media, news headlines specifically are important in the structure of newspaper articles as there is an assumption by readers that "headlines express the most important topic of the news" (Dijk 1985). If headlines express the most important topic of the news, then what do the headlines about ChatGPT and artificial intelligence reveal?

In 2016, Ethan Fast and Eric Horvitz analyzed thirty years of articles, from the New York Time, related to artificial intelligence to find the long-term trends through sentiment analysis and the ideas associated with artificial intelligence through word frequency analysis. They found that over thirty years, the sentiment of the articles remained consistently two to three times more positive, focusing on technological innovations, entrepreneurship and market growth rather than having negative sentiments. In terms of specific negative issues, they found a clear upward trend over time of specific issues such as "loss of control", "ethical concerns", and "negative impact on work" (Fast & Horvitz 2016). Has the sentiment changed since the launch of ChatGPT and what are the key issues being discussed since the launch of ChatGPT in the mainstream media headlines? How does the mainstream media currently portray negative issues such as AI's impact on jobs, ethical considerations, bias, diversity and copyright concerns?

The below two headlines present the polarities in the way ChatGPT and artificial intelligence is portrayed in the mainstream media. While both headlines discuss Big Tech, the first focus on the negative in terms of power relations and impact on job, while the other celebrates the gold rush of investment funds into artificial intelligence in Silicon Valley (Chowdhury 2023, Tiku 2023).

> *"Tech firms hold all the power right now as Goldman Sachs predicts AI will impact 300 million jobs"*
> *Business Insider (Chowdhury 2023)*

> *"AI is reviving San Francisco's tech scene. Welcome to 'Cerebral Valley.'" Washington Post (Tiku 2023)*

This article presents a quantitative data analysis of the early trends revealed by analyzing word frequency and sentiments of a corpus of mainstream news headlines in regards to the subject of artificial intelligence since the launch of ChatGPT in November 2022 to March 2023. This author is interested in analyzing how frequently has ChatGPT and artificial intelligence been discussed in the media and how the number of occurrences has increased in frequency since its launch. What the most frequent words in news headlines about ChatGPT and artificial intelligence and what do they reveal in terms of agenda-setting of mainstream media? What issues and actors are raised in the headlines, and which issues and actors are represented poorly or are absent from the corpus, which, as according to Fast & Horvitz (2016) such as jobs and ethics were of concern to the public? What happens when we investigate Diversity, Equity, and Inclusion terms in newspaper headlines related to ChatGPT related and Artificial Intelligence? To what extant can the distant reading of the corpus, such as sentiment analysis scores, reveal trends in how the mainstream media is addressing the issues to the public, about ChatGPT and artificial intelligence, and can this method be relied upon when compared to closer reading methods? Do the headlines reveal a bias in the media towards the topic of ChatGPT and what can be revealed from the analysis in relation to the connections between Big Tech, Big Media and Big Data?

This article contributes to the growing discourse to inform the agenda of mainstream media in what it chooses to portray about ChatGPT and artificial intelligence to the public and what as issues and actors dominates the trends and which are absent, and what do these absences reflect on a critical level.

## II. METHODS

This is a mixed-method study conducted on 10,902 news headlines collected from Western media sources predominately located in the United States. The New York Times, Wall Street Journal, and New York Post are examples of newspapers which are considered mainstream media publications in the United States and are part of the corpus (Sherer & Mitchell 2021). First, the total amount of headlines occurring per month was counted to analyze if there was an increase in interest in the mainstream media about the topic of ChatGPT reflective of ChatGPT's user acquisition scale. Then word frequency analysis and sentiment analysis was conducted on the corpus. Next, manual sentiment analysis was conducted on a sample of the corpus to qualitatively compare and contrast the sentiment results. This section presents each of the steps of the research methodology starting with data collection, which is then followed by data preprocessing, and analysis.

### A. Data Gathering

The launch of ChatGPT to the public, by its parent company OpenAI, in late November 2022, was used as the starting point of the collection of the headline corpus till the 29th of March 2023. Two data gathering techniques were employed by this author. The first data gathering technique was to manually collect the news per day from Google News, which is a popular news aggregator. While collecting the data through this manual technique, it was found that Google News only lists a maximum of twenty articles in its front facing archives per day for dates older than two months, and only provides full access to about 150 daily for recent dates. 1087 news headlines were collected manually. Due these issues, this data gathering technique was disregarded as it would have resulted in an imbalanced viewpoint in terms of dates and articles, as only dates in the past two months would have multiple articles and anything before that date would be limited to a maximum of twenty articles per day. To have a more balanced analysis another data collection method was required to collect a corpus large enough to perform the text mining analysis.

For the purposes of this article, the Perigon API was used as it complies with copyright law and provides a freemium model. The Perigon company is based in Texas. Perigon, is a news aggregator company based in Texas which delivers news from over 70,000 sources on the internet(Perigon 2023. Pergon complies with the Digital Millennium Copyright Act (DMCA), which is a United States federal law which protects copyright issues due to new technology (Perigon 2023, LLI 2022). Perigon's newspaper sources include: The Washington Post, The New York Times, The Guardian, The New Yorker, Wired, TechCrunch, Forbes and The Wall Street Journal as well as news aggregators such as Yahoo (Perigon 2023). While Perigon is a paid news aggregator service focused on corporations as clients searching to find trends and sentiments in the news related to their businesses, Perigon also offers a freemium model for developers and provides them with an API key to their database, within a certain limited data range. The Perigon API platform allows for various customized search options when retrieving the data required such as topic, date, news headline or article text, keywords, and type of news outlet when retrieving the data. The data is then provided in a JSON (JavaScript Object Notation) format. This author contacted Perigon, through email, and was informed that the data required would be included as part of their freemium model as it was within the scope of the data range Perigon had set for developers.

The search parameters, used to collect the dataset, started on the 23rd of November 2022, ten days before ChatGPT was launched, and extended to the 29th of March 2023. The language was set to English for ease of analysis in terms of quantitative analysis and NLP capabilities of this author. The source/media was selected as Top 100, which can be chosen as the most popular sources globally as set by Perigon's algorithms which includes most of the sources listed earlier, and are primarily located in the United States and Europe. The selected keywords were: AI, Artificial Intelligence, and ChatGPT. Since the public launch of ChatGPT3 in November 2023, ChatGPT has become synonymous with AI and the general public use these terms interchangeably (Saha 2023). The resulting corpus included 10,902 newspaper headlines with their respective URLs and publishing date. The dataset was provided in JSON, by Perigon, and then converted to a CSV file.

The research findings are limited and do not show a full global multilingual view of the topic but rather a predominantly Global North viewpoint of the topic with an emphasis on popular newspaper outlets in the United States. This is not only due to the English language considerations but also to the selection process of Perigon of its Top 100 Media/Sources, which is another opaque layer to this author. Since Perigon is a private company, the source code of how its algorithms and aggregating processes are accomplished are also opaque to this author. To add a layer of verification to the resulting corpus which was used in this article, this author used the manually collected newspaper headlines to manually crosscheck the Perigon dataset, and to make sure it was accurate with 100 headline samples, URL links and various publishing dates.

*B. Data Preprocessing*

Each headline, and related date and URL link was preprocessed to prepare the corpus for natural language processing (NLP) tasks including frequency-sorted word lists, sentiment analysis and topic modeling. Preprocessing the data is important to clean and reduce the noise in the data and help with performance of the classification process for topic modeling and sentiment analysis (Haddi et al. 2013). This author used the open source platform Anaconda which allows the use of Python, an open source programming language used for "for large-scale data processing, predictive analytics, and scientific computing" (Kadiyala 2017). In addition, filtered dataset functions were used in Microsoft Excel to provide specific keyword analysis. This author acted as the prompt engineer for the ChatGPT3 Plus and ChatGPT4 chatbot (OpenAI 2023), which wrote all the python programming code for this article and which was then executed in the Jupyter file.

This author used the Natural Language Toolkit (NLTK) for preprocessing the corpus in addition to Microsoft Excel functions. NLTK is "a suite of open-source program modules providing ready-to-use computational linguistics courseware…covering statistical natural language processing, and is interfaced to annotated corpora" (Loper & Bird 2002). The following preprocessing techniques were applied to each newspaper headline using NLTK: removal of duplicates headlines, headlines using keywords, artificial and intelligence, but which are not related to AI, tokenization, removal of non-ASCII characters, lemmatization, removal of punctuation, lowercasing all letters, and removal of incomplete headlines ending with three dots. Removal of stopwords was done were needed in specific NLP and Text Mining Techniques. This process reduced the corpus from 10,902 headlines to 7,989 headlines.

In the corpus, each headline is linked to a specific date of publication and URL link. Each date of publication was made up of ISO 8610 timestamp with date, time, timezone offset. The time and timezone offset were split from the date, and the date string converted to integers. For example: 2022-11-23T08:12:05+00:00 was transformed to 11/23/2022 which allowed it to be used as a proper date in data visualizations. To create a proper name list of publications and aggregators, the URL domain name was split from the protocol, path and query parameters, leaving the publication or aggregator name. The list of one hundred publication or aggregator names was manually cleaned using Microsoft Excel's find and replace function to insert white spaces and to write full publication name.

*C. NLP and Text Mining Techniques*

The following NLP and text mining techniques were conducted on the corpus: frequency-sorted word lists and sentiment analysis.

Frequency-sorted word lists have historically been used as a method to understanding corpora (Baron et al. 2009). The frequency-sorted word lists register how many times a word occurs in the corpus. This technique was applied to the corpus as it provides insight on the presence of the most frequent words related to ChatGPT related headlines, and to provide indications of absent or low frequency words. The sentiment scores were appended to the corpus in a CVS file, and then further analyzed using pandas and NumPy which are data analysis python libraries, and Matplotlib and Seaborn which are python toolkits for data visualizations (McKinney 2010, Harris 2020, Hunter 2007, Waskom 2021).

Sentiment analysis has been used for in many different fields including sociology, marketing, and advertising (Hutto & Gilbert 2014). Sentiment analysis was conducted on the news headlines in the corpus to potentially investigate the different sentiments which occur in the headlines and how they could be mapped out over time in terms of positive, negative and neutral outputs. News headlines are generally short in terms of word count are on average of 12 words in the corpus, and are similar in scale to tweets and social media output. This author used Vader, an open-source sentiment analysis tool in the NLTK toolkit, as Vader was developed specifically for social media tweets and micro-blogging instead of larger scale articles and texts (Hutto & Gilbert 2014). Vader has also been shown to perform " as well as individual human raters at matching ground truth" (Hutto & Gilbert 2014). Each headline in the corpus was classified as positive, negative or neutral. The sentiment scores were appended to the corpus in a CVS file, and then further analyzed using then further analyzed using pandas and NumPy, and Matplotlib and Seaborn for data visualizations (McKinney 2010, Harris 2020, Hunter 2007, Waskom 2021).

This allowed for the visualization of how the sentiment evolved over time, in a stacked bar chart. Data visualizations of the sentiment scores of specific keywords such as "job",

"ethic" and "woman" were developed to compare and contrast the total sentiment frequency changes across news headlines in terms of the full corpus versus when discussing one specific keyword in the corpus. This author conducted a manual close reading of all the sentiments scores of headlines related to the keyword: job, to compare and contrast with the results of the Vader sentiment scores. All data is available.

## III. RESULTS

*A. Headlines per month*

In Figure 1 the total amount of headlines occurring per month, a bar plot was plotted in Figure 1. There is a sharp increase in the number of news headlines related to ChatGPT and artificial intelligence from December 2022 to March 2023. This indicates that the trending interest of mainstream media in regards to the topic of ChatGPT and AI. This increase was also related to OpenAI's release of ChatGPT at the end of November and that is indicated in the low number of headlines in November of 173 occurrences as data was only collected from the 23rd of November 2022. In the subsequent months of December 2022, January 2023 and February 2023, the number of headlines approximately doubled from month to month rising from 748 in December to 2716 occurrences in February. The trend of increasing number of articles per months can also be attributed to the continuous launch of AI products within this small timeframe. In early February, Microsoft relaunched their search engine Bing integrated with OpenAI's ChatGPT 4, a result the one-billion-dollar investment by Microsoft in OpenAI (Lardinois 2023). In March, OpenAI also released their new ChatGPT model, ChatGPT4, which according to OpenAI "exhibits human-level performance on various professional and academic benchmarks, including passing a simulated bar exam with a score around the top 10% of test takers" while ChatGPT3 passed the bar exam in the bottom 10% of test takers (OpenAI 2023). On the 21st of March 2023, Google released in limited capacity its AI chatbot Bard to compete with OpenAI and Microsoft (Alba & Love 2023).

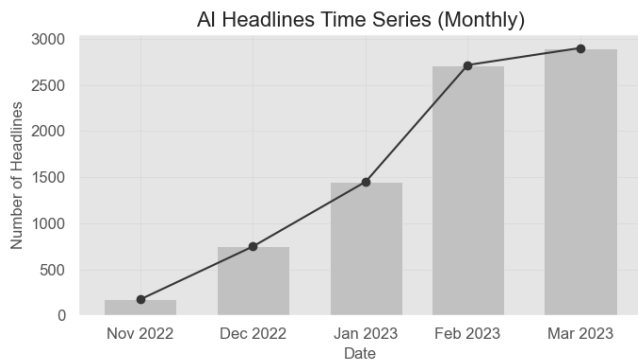

Figure 1: Bar plot of news headline occurrences per month

*B. Top 30 Word Frequency Rankings and Percentages*

Table 1 and Figure 2 represent the top thirty frequently mentioned words with their rankings in the headline corpus. The research question of this study is to investigate what topics are represented, underrepresented and what is absent in the news headline about ChatGPT and AI. Thus keywords were added to Table 1 and Figure 1 to track the a variety of topics including ethics, copyright, bias, and DEI (Diversity, Equity, and Inclusion) in terms of representation in the news headlines about ChatGPT and artificial intelligence. The following keywords were added: "race", "ethic", "bias", "copyright", "woman", "equity", "diversity", "inclusion" and "gender".

| Rank | Keywords | Frequency |
| --- | --- | --- |
| 1 | ai | 5401 |
| 2 | chatgpt | 2041 |
| 3 | new | 703 |
| 4 | artifical | 641 |
| 5 | intelligence | 641 |
| 6 | google | 476 |
| 7 | microsoft | 441 |
| 8 | stock | 396 |
| 9 | technology | 357 |
| 10 | generative | 351 |
| 11 | launch | 351 |
| 12 | market | 343 |
| 13 | chatbot | 330 |
| 14 | tech | 289 |
| 15 | aipowered | 288 |
| 16 | company | 265 |
| 17 | platform | 255 |
| 18 | tool | 253 |
| 19 | announces | 226 |
| 20 | openai | 216 |
| 21 | post | 208 |
| 22 | council | 193 |
| 23 | help | 192 |
| 24 | solution | 190 |
| 25 | business | 178 |
| 26 | bing | 178 |
| 27 | make | 176 |
| 28 | job | 172 |
| 29 | data | 171 |
| 30 | startup | 169 |
| 31 | race | 119 |
| 32 | ethic | 52 |
| 33 | bias | 36 |
| 34 | copyright | 32 |
| 35 | woman | 31 |
| 36 | equity | 10 |
| 37 | diversity | 8 |
| 38 | inclusion | 4 |
| 39 | gender | 1 |

Table 1: Top thirty word frequency table with additional keywords

These additional keywords are limited in this article, and represented a fair 30% additional words added to the ranking. A wider sample of additional keywords may reveal more inclusions related to the represented, underrepresented and what is absent in the news headline about ChatGPT and AI but this is currently outside the scope of this article.

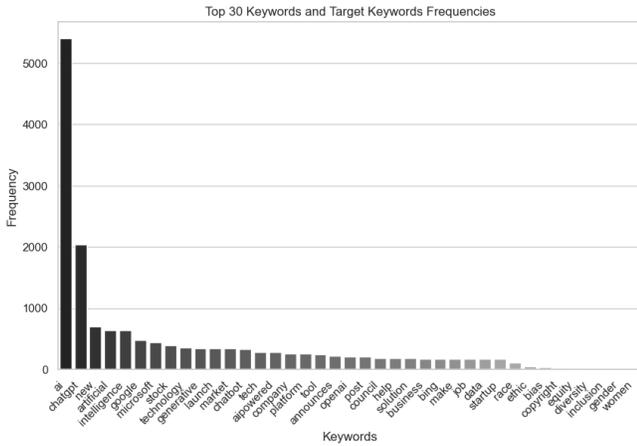

Figure 2: Bar plot of Top thirty word frequency table with additional keywords

In the top five word frequency, the words "ai", "chatgpt", "artificial" and "intelligence" were featured due to the fact that they were the keywords inserted as search parameters and had a cumulative word frequency of 8724 mentions. Big Tech companies such as Google and Microsoft were ranked in the top ten, while OpenAI, the company that developed ChatGPT, was ranked at twenty. Google, Microsoft, Bing and OpenAI had a cumulative word frequency of 1311 mentions and represented 17%[i]. of the total mentions. Technology related keywords were featured heavily in the top thirty list including: "technology", "generative", "chatbot", "aipowered", "platform", and "data" had a cumulative word frequency of 2463 mentions which represented 32% of the total mentions. Keywords, related to the launch of ChatGPT as a product, were also featured heavily and included the words: "new", "launch" and "startup". They had a cumulative word frequency of 1223 mentions which represented 16% of the total mentions.

The keyword "job" was at the bottom of the top thirty word frequencies with a ranking of twenty-eight and a cumulative word frequency of 172 which represented 2% of total mentions. The additional keywords of "ethic", "bias", "copyright", "woman", "equity", "diversity", "inclusion", "black" and "gender" were ranked the lowest because they were added keywords and computationally should not be present in the visualizations of the top thirty rankings. The additional keywords have a cumulative word frequency of 263 mentions which represented 4% of total mentions.

Overall Big Tech companies and the launch of ChatGPT dominated the top thirty word frequency rankings with an overall 65% of the top mentioned word frequencies while the additional keywords and jobs, which the research questions of this article is analyzing in terms of keyword absence or low representation in the corpus, only represented 6% of the mentions.

*C. Sentiment Analysis*

Table 2 provides the cumulative sentiment distribution into negative, neutral and positive sentiment labels. The majority of the headlines had a positive or neutral sentiment, with 6,486 instances, representing 81% of the headlines, while the negative sentiment, with 1,503 instances represented 19% of the headlines. This indicates that the mainstream media tends to have dominant positive and neutral sentiments in regards to ChatGPT and AI. Table 3 provides sample headlines with different sentiment labels.

| Sentiment | Number of Headlines |
|---|---|
| Positive | 3427 |
| Neutral | 3059 |
| Negative | 1503 |

Table 2: Sentiment distribution of all headlines

| Headline | Sentiment |
|---|---|
| Mind-blowing new AI chatbot writes sophisticated essays and complicated coding | Positive |
| "Like We Just Split The Atom": ChatGPT AI Shakes Up Tech | Positive |
| Explainer: ChatGPT - what is OpenAI's chatbot and what is it used for? | Neutral |
| Bank of America sees AI transforming the internet over the next five years—top stocks to ride trend | Neutral |
| ChatGPT, an AI chatbot, has gone viral. Some say its better than google; others worry its problematic | Negative |
| Lawsuit takes aim at the way AI is built | Negative |

Table 3: Sample headlines with different sentiment labels

To study the sentiment trends over time in correlation to the total amount of headlines occurring per month, a stacked bar plot was plotted, alongside their respective sentiment percentages of positive, natural and negative headlines in Figure 3. Over the four-month period analyzed, the total number of news headlines with negative sentiment was 1503 occurrences and were 31% of the total number of headlines with positive sentiment with 3427 occurrences.

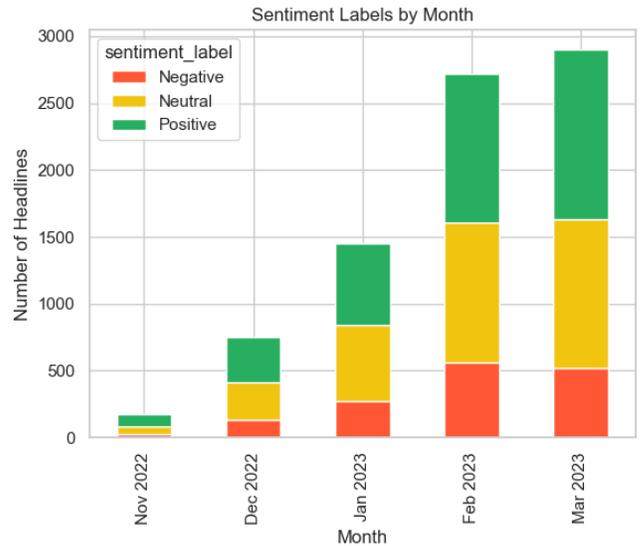

Figure 3: Sentiment trends over time

While the negative sentiment totals are approximately 56% lower than the positive sentiment totals, a closer reading into specific key words, such as "job" and "bias" and "diversity" reveal a different result to the top heavy positive and neutral sentiment totals in the overall corpus. Figure 4 demonstrates the sentiment totals for headlines which include the keyword "job". Unlike the total sentiment totals, the negative sentiment total is now higher than both the positive and neutral, revealing that in regards to specific keywords such as "job" the sentiment in the mainstream media becomes more negative as a trend in relation to artificial intelligence.

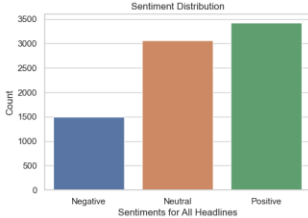
Figure 4: Total sentiment scores all headlines

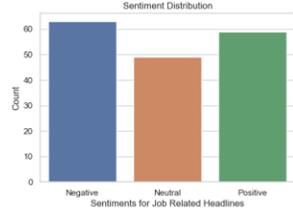
Figure 5: Total sentiment scores for keyword "job" related headlines

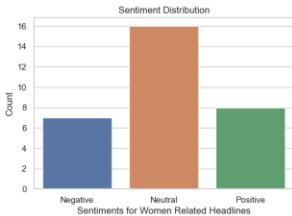
Figure 6: Total sentiment scores for keyword "woman" related headlines

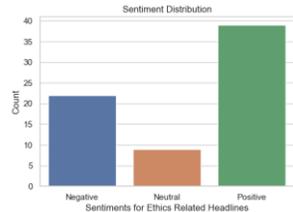
Figure 7: Total sentiment scores for keyword "ethic" related headlines

In Figure 6 the sentiment scores for keywords scores, where the keyword "woman" was present in the headlines revealed a nearly equal negative and positive total sentiment score, even though the negative sentiment was slightly lower than the positive but still trending differently from total headlines sentiment. In Figure 7 the keyword "ethic" demonstrated a very high level of positive sentiments.

Due to low level of occurrences of the keywords "job", this author chose to analyze manually on of the keywords "job" headlines to give them a sentiment from the positionality of this author who is a interdisciplinary feminist researcher. This exercise was done to compare and contrast the results of Vader's sentiment labels and this human researcher's sentiment scores from their positionality and to reflect on the scientific methodology of using Vader sentient analysis. Table 4 provides sample headlines with the Vader sentiment score and the human sentiment score.

| Headline | Vader Sentiment Score | Human Sentiment Score |
|---|---|---|
| The Future Of AI Is Finally Here – And A Lot Of People Are Going To Be Out Of A Job | Neutral | Negative |
| The owner of Insider and Politico tells journalists: AI is coming for your jobs | Neutral | Negative |
| Meet ChatGPT, the artificial intelligence chat bot that could take your job | Positive | Negative |
| We asked ChatGPT which jobs it thinks it will replace—and it's not good news for data entry professionals or reporters | Positive | Negative |

Table 4: Sample headlines with Vader label results and the Human label results

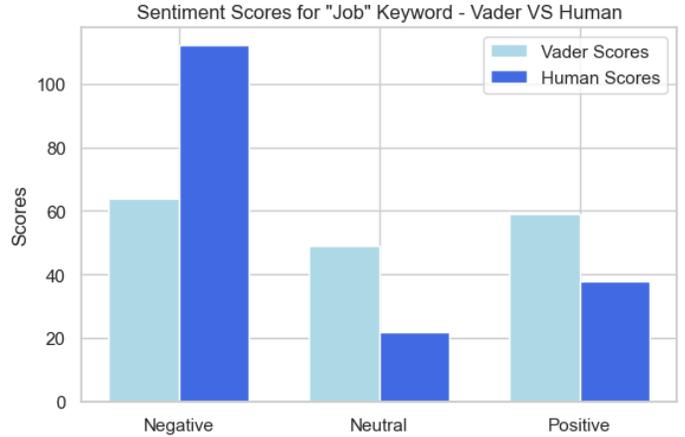
Figure 8: Sample headlines with Vader label results and the Human label results

In Figure 8, the results reflected that a human sentiment labeler had nearly double the negative sentiment labels of the Vader sentiment label score, while the neutral and positive also had lower percentages of occurrences. This result pushes against the positive and natural sentiments which dominate the overall corpus sentiment scores in Figure 4. While it is out of the scope of this article to analyse the Vader sentiment analysis as a methodological tool, Figure 6, reflecting the keyword "job" sentiment does align more with the previous study of Fast and Horvitz (2016) and highlights the need for close reading of the headlines to fully understand their sentiment and what they reveal in terms of trends.

## IV. DISCUSSION

OpenAI, based in Silicon Valley, was cofounded by billionaire Elon Musk, the cofounder of Tesla and Twitter, and billionaire Peter Thiel, co-founder of PayPal; in addition, Microsoft, cofounded by Bill Gates, invested over one billion dollars recently into OpenAI (Kay 2023). OpenAI, as a company, is surrounding by the presence of Big Tech actors. Big Tech and the actors surrounding it dominated the word frequency results found in the corpus mainstream media headlines, with an overall representation of 65% with a high score of positive and neutral sentiment scores. Since Big Tech is predominately dominated by white males (Daileda 2016), the overall representation in the corpus slants to a hegemonic outlook of the mainstream media and mainly represents the wealth, white, Western, capitalist male figure as the main protagonist in the corpus. The etymology of the word "hegemony" is derived from the Greek term "hēgemonia" which means "to have dominance over" and is used to describe the relationships between different subjects. Raymond Williams, a cultural theorist, argues that hegemony is an active process and is constantly in motion, adapting and adjusting "values, meanings and practices" as needed by the dominant ruling class to constantly preserve their power (1977). These "values, meaning and practices" are merged into a culture, to become the dominant social order. In the case of ChatGPT, Big Tech and its actors reflect these dominant groups and the mainstream media, which are also generally also all owned by large conglomerates. This reflection of capitalism as an ideology which favors the few over the majority is reflected in the results of this article which was one of its objectives.

In her seminal essay, Situated Knowledges: The science question in feminism and the privilege of partial perspective, Donna Haraway defines the god trick as the act "of seeing everything from nowhere, but to have put the myth into ordinary practice" (1988). She argues the history of science goes hand in hand with capitalistic ideology and male supremacy which reflects the interests of the powerful and who keep all others at a distance to maintain power. By mobilizing Haraway's argument onto the corpus, the corpus becomes a "god trick" in and of itself. The corpus was collected through a black box API, from a private company, it is reflective of only mainstream newspapers, which are owned by large capitalist conglomerates, and the results conclude that the majority of the news is positive or neutral in regards to Big Tech projects and actors who dominate 65% of the content of the corpus. From the viewpoint of agenda-setting theory, the word frequency results also reveal the mainstream medias insistent agenda promoting the mythology of technological progress

Goldman Sachs predicted that over 300 million jobs would be lost or effected by AI (Kelly 2023). In their article, An Early Look at the Labor Market Impact Potential of Large Language Models, OpenAI findings revealed "that around 80% of the U.S. workforce could have at least 10% of their work tasks affected by the introduction of LLMs, while approximately 19% of workers may see at least 50% of their tasks impacted". This large-scale impact on jobs, as per Goldman Sachs and OpenAI, was reflected in the only 2% of the corpus with 172 mentions, with a positive and negative sentiment of nearly the same values, and it these numbers are to be believed, then mainstream media with its agenda-setting does not think people's job and the loss of job functions due to artificial indolence is a topic worth pursuing, as it does not align with the positive mythology of technological progress. The manual closer reading of the "job" keyword, further exposed the inaccuracies underling the scientific methodology provided by the distant reading of the corpus and that the newspaper headline corpus needed to be critically assessed with further close readings. The anxiety of job loss due to new technologies brings the Luddite revolution into focus.

Colloquially Luddites are known as technophobes, wary of technology but the Luddite revolution, which took place during the Industrial revolution, was actually one about the impacts the new technology of the mechanized looms had on worker's job security.  The Luddites were not protesting "against machines" but rather against the uncalculated social and monetary costs the loss of jobs the new machine bought with it, while it accumulated wealth for the factory owners (Benjamin 37). Lord Byron, who was the father of Ada Lovelace, one of the first computer programmers on the cusp of another technological evolution, defended the Luddites in the House of Lords. He said "it cannot be denied that they have arisen from circumstances of the most unparalleled distress …. nothing but absolute want could have driven a large, and once honest and industrious, body of the people, into the commission of excesses so hazardous to themselves, their families, and the community" (Pirie 2019). Similar to the anxieties of the Luddites, the technological progress of AI will bring with it an impact to jobs which is enormous according to Goldman Sachs and Open AI, thus bringing forth the question of what agenda-setting is being structured when an issue, such as jobs, which will affect over 50% of the work force is not being addressed in the mainstream media.

In their book, Data Feminism, Catherine D'Ignazio and Lauren F. Klein argue that to remake the world there is a need to make each data project an analysis of how "power operates in the world" (21).  They argue that that in each data project, the matrix of domination, a theory developed by Patricia Hill Collins, "works to uphold the undue privilege of dominant groups…who hold more economic, social and political power… while unfairly oppressing minoritized groups" (24). By mobilizing D'Ignazio and Klein's theory of each data project being seen as power analysis project which should asking who benefits from the data, and whose goals are prioritized (26), the analysis of the corpus of mainstream headlines about ChatGPT and artificial intelligence becomes a story of power and a story of absence.

In the corpus, where are the concerns of ChatGPT and artificial intelligence in terms of impact on the job market? Where is the "Other" located in the corpus, in terms of inclusion of minorities, diversity, equity and inclusivity issues? Where does it leave the ethical issues of copyright and bias? The results of the analysis of the corpus reveal the hegemony of the corpus and its privileging of dominant groups of Big Tech and the capitalist system. The corpus of mainstream headlines allocated only 6% of the corpus to the following keywords: jobs, ethics, bias, copyright, woman, equity, diversity, inclusion, black and gender. Donna Haraway argues that "it matters what stories we tell to tell other stories with" (12, 2016). If one is to agree with Haraway, the poorly represented absent stories of the minoritized groups are the real results of this article.  These absent stories and who they represent do not fall into the agenda-setting of Big Tech and Big Media, and do not align with their priorities and thus become stories of absent data never to be narrated. In the case of the mainstream newspaper corpus, each of the headlines tell a story, and predominately the story this corpus tells is on the collusion between Big Tech and Big Media in agenda-setting.

V. RELATED WORKS

Due to the limited time frame since ChatGPT's launch in November 2022, a limited number of articles has been published in regards to the public perception of ChatGPT. The two most closely related works are: ChatGPT: A Meta-Analysis after 2.5 Months by Leiter et al. (2023) and "I think this is the most disruptive technology" Exploring Sentiments of ChatGPT Early Adopters using Twitter Data by Haque et al

(2022). Haque et al. (2022) focused on analyzing the sentiment of tweets of ChatGPT early adopters and topic modelling. They found that 83% of tweets were positive in sentiment about ChatGPT, similar to the dominant positive sentiment found in this article from the mainstream headlines. Leiter et al. (2023) analyzed 300,000 tweets, in multiple languages, and over 150 scientific papers about ChatGPT. They found that overall ChatGPT tweets had positive sentiments in the English language, but more negative sentiments where found in other languages. While the datasets differed from that of this article, positivity as a sentiment seems to be a common thread between all including the thirty-year analysis by Fast and Horvitz (2016).

## VI. Conclusion

The exponential growth in user acquisition and popularity of OpenAI's ChatGPT, an artificial intelligence(AI) powered chatbot, was accompanied by widespread mainstream media coverage. This article presents a quantitative data analysis of the trends and sentiments revealed by conducting text mining and NLP methods onto a corpus of 10,902 mainstream news headlines related to the subject of ChatGPT and artificial intelligence, since the launch of ChatGPT3. In the sentiment analysis, ChatGPT and artificial intelligence were perceived more positively than negatively in the mainstream media. In regards to word frequency results, over sixty-five percent of the top frequency words were focused on Big Tech issues and actors while topics such as jobs, diversity, ethics, copyright, gender and women were poorly represented or completely absent and only accounted for six percent of the total corpus. This article is a critical analysis into the power structures and collusions between Big Tech and Big Media in their matrix of domination and exclusion of the key issues of diversity and inclusion from mainstream media.

## Notes

Percentage calculations note: Search parameters keywords: AI, artificial, intelligence and ChatGPT were removed from the percentage calculation presented as they are overly represented due to the fact that they were the initial search parameters and represented by themselves 8724 out of the 16384 mentions which is over 50% of total mentions. Percentages were rounded to the nearest integer.